\documentclass[aps,floatfix,showpacs,showkeys]{revtex4}
\usepackage{color,graphicx}
\usepackage{dcolumn}
\usepackage{bm}
\usepackage{epsfig}
\usepackage{supertabular}
\usepackage[bookmarksopen]{hyperref}
\def\be {\begin{equation}}
\def\ee {\end{equation}}
\def\nn {\nonumber}
\def\bea {\begin{eqnarray}}
\def\eea {\end{eqnarray}}

\begin{document}
 \title{Wake in anisotropic Quark-Gluon Plasma }
\bigskip
\bigskip
\author{ Mahatsab Mandal}
\email{mahatsab.mandal@saha.ac.in}
\author{ Pradip Roy}
\email{pradipk.roy@saha.ac.in}
\affiliation{Saha Institute of Nuclear Physics, 1/AF Bidhannagar
Kolkata - 700064, India}

\begin{abstract}
We calculate the wake in charge density and the wake potential due to the passage of a fast parton in an 
anisotropic quark gluon plasma(AQGP). For the sake of simplicity small $\xi$(anisotropic parameter) limit has 
been considered. 
When the velocity($v$) of the jet is parallel to the anisotropy
direction(${\hat n}$) and remains below the phase
velocity($v_p$), the wake in induced charge density shows a little oscillatory behavior in AQGP, contrary to 
the isotropic case. With the jet velocity greater than the phase velocity, the oscillatory behavior 
increases with $\xi$. Also for $v>v_p$ one observes a clear modification of the cone-like structure in presence 
of anisotropy. For the parallel direction in the backward region, the depth of the wake potential decreases with the 
increase of $\xi$ for $v<v_p$ and the potential becomes modified Coulomb-like for higher values of $\xi$. 
In the forward region, the potential remains modified Coulomb-like with the change in magnitude for nonzero $\xi$, 
 for both $v>v_p$ and $v<v_p$. In the perpendicular direction, the wake potential is symmetric in the forward and 
backward regions. With the increase of $\xi$, the depth of negative minimum is moving away from the origin 
irrespective of the jet velocity. 
On the other hand when the jet velocity
is perpendicular to the anisotropy direction, we find significant changes
 in the case of both wake charge density and potential in comparison to
the isotropic case. For nonzero $\xi$, the oscillatory nature of the 
color charge wake is reduced at $v>v_p$. Also the oscillatory 
behavior of the wake potential along the direction of motion of the parton
is attenuated in the backward direction for anisotropic plasma at parton
velocity $v>v_p$. In the presence of anisotropy, for $v<v_p$, the screening 
potential along the perpendicular direction of the parton is transformed from
the Lennard-Jones type to a modified Coulomb-like potential.

\end{abstract}
\keywords{wake, quark-gluon plasma, anisotropy}
\pacs{25.75.-q, 12.38.Mh}

\maketitle

\section{Introduction}
One of the primary goals of ultrarelativistic heavy-ion collision experiments at 
BNL RHIC and at CERN LHC is to create and explore the novel state of matter in which 
quarks and gluons are deconfined, commonly known as quark-gluon plasma(QGP).
In the early stage of heavy-ion collision, high $p_T$ partons produced by the hard scatterings 
will travel through the hot and dense medium and lose energy, mainly by radiative processes.
As a consequence high $p_T$ hadrons produced due to parton fragmentation are suppressed.  
This phenomenon is the so-called jet quenching, because in the direction of propagation of the jet
one observes a decrease of high energy hadrons and increase in the soft hadrons
~\cite{jetquen1,jetquen2,jetquen3,jetquen4,jetquen5,jetquen6,jetquen7,jetquen8,jetquen9}. 
Moreover the experimental azimuthal dihadron distribution at RHIC shows a double peak structure 
in the away side~\cite{prl95,prl97} for the intermediate $p_T$ particles. 
The wake induced by the jets are proposed as one of the possible
explanations for the formation of double peak structure in two-particle correlation~\cite{pg_34,yy}.
However, recent studies show that the observation of double peak
structure 
could be because of initial
energy density fluctuations as has been argued in 
Refs.~\cite{prl103,prc86}. 
Nevertheless, the passage of jets through the plasma 
leads to the formation 
of wakes~\cite{plb618,prd74,jpg34,npa856,jpg39}, Mach cones~\cite{npa750,jconf,pg_34} and 
Cerenkov radiation~\cite{prl96,prc73,npa767}, which may be observable as collective excitation 
and shock wave in heavy-ion collisions.

When a test charge moves in the plasma, it creates a screening potential. The screening 
potential for a heavy quark-antiquark pair is strongly anisotropic and loses forward-backward symmetry with 
respect to the direction of the charged particle~\cite{prd39}. First, Ruppert and Muller~\cite{plb618} have
investigated that when a jet propagates through the medium, a wake of current and charge density 
is induced which can be studied within the framework of linear response theory in two different scenarios: 
(i) a weakly coupled quark-gluon plasma described by hard thermal loop(HTL) perturbation theory and 
(ii) a strongly coupled QGP, which has the properties of a quantum liquid. 
The result shows the wake in both the induced charge and current density due to the screening effect of the 
moving parton. On the other hand, in the quantum liquid scenario, the wake exhibits a oscillatory behavior when 
the charge parton moves very fast. Later, Chakraborty et al.~\cite{prd74}  also found the oscillatory behavior of 
the induced charge wake in the backward direction at large parton speed using the high temperature approximation. 
In a collisional quark-gluon plasma, it is observed that the wake properties change significantly compared to the 
collisionless case~\cite{jpg34}. Recently, Jiang et al~\cite{npa856,jpg39} have investigated the color response 
wake in the viscous QGP with the HTL resummation technique. It is shown that the increase of the shear viscosity 
enhances the oscillation of the induced charge density as well as the wake potential.

In all the above phenomenological treatments of the plasma the system is assumed to be isotropic in momentum space. 
However, in the very early stage of heavy-ion collision, due to rapid longitudinal expansion of the matter
along the beam axis the system becomes colder along the longitudinal direction than the transverse direction, 
leading to $\langle p_L^2\rangle<\langle p_T^2\rangle$. Consequently, the plasma is anisotropic in momentum space.
To characterize this anisotropy many signals have been proposed~\cite{prd68,prl100,prc78,prc79,prc81,prc83,prc84}. 
We, in this work, will study the wake induced by a jet propagating through anisotropic quark-gluon plasma.
The dielectric response function contains all the information of the chromoelectromagnetic properties of 
the plasma. In the presence of momentum-space anisotropy, the distribution functions of the quark and gluon 
are modified and it will affect the dielectric function. Moreover, due to anisotropic momentum distribution 
at very early stage of the heavy-ion collision, the collective modes of the AQGP may be unstable~\cite{prd68,prd70}. 
These unstable modes  grow exponentially in time, which leads to a more rapid thermalization and 
isotropization of the soft modes in QGP. Such process may play a significant role in the dynamical properties 
of the QGP. For example, the induced charge density and wake potential will be influenced by the 
anisotropic effect of the plasma as we shall see in the following. 

The current work is organized as follows: In section 2 we briefly recall the necessary ingredients to calculate
the induced color charge density and screening potential in QGP. Section 3 is
devoted for the evaluation of dielectric tensor in AQGP. In section 4 and 5 the wake in charge density and potential
will be described along with the numerical results. Finally, we conclude in section 6.

\section{Basic Equations}
The covariant form of the Maxwell equation is 
\be
\partial_{\mu}F^{\mu\nu}(K)=J^{\nu}_{ind}(K)+J^{\nu}_{ext}(K),\label{max} 
\ee
where the total current is composed of the induced current $J^{\nu}_{ind}$ 
and the external current $J^{\nu}_{ext}$ introduced from the external
sources. In the linear approximation, the equation of the gauge field 
is represented by the induced current in terms of the self-energy
~\cite{prd62,prd68}:
\be
J^{\mu}_{ind}(K)=\Pi^{\mu\nu}(K)A_{\nu}(K)\label{ind}.
\ee
where according to the semiclassical transport theory, the hard thermal loop resummed  gluon self-energy can be 
written as~\cite{prd68} 
\be
\Pi^{\mu\nu}(K)=g^2\int \frac{d^3p}{(2\pi)^3} P^{\mu}
\frac{\partial f({\bf p})}{\partial P_{\beta}}\Big(g^{\beta\nu}-\frac{P^{\nu}K^{\beta}}{K.P+i\varepsilon}\Big)
\ee
One can easily show that the polarization tensor is symmetric, $\Pi^{\mu\nu}(K)=\Pi^{\nu\mu}$ and transverse,
$K^{\mu}\Pi^{\mu\nu}(K)=0$. In deriving the above equation it is assumed that the momentum is soft, 
$k\sim~gT\ll T$ and the magnitude of the field fluctuations is $A\sim\sqrt{g} T$.
By combining Eqs. (\ref{max}) and (\ref{ind}) the external current can
be related to the gauge field as 
\be
[K^2{\rm g}^{\mu\nu}-K^{\mu}K^{\nu}+\Pi^{\mu\nu}(K)]A_{\nu}(K)=-J^{\mu}_{ext}(K).
\ee
In the temporal axial gauge, where $A_0=0$, the above equation becomes,
\bea
[(k^2-\omega^2)\delta^{ij}-k^ik^j+\Pi^{ij}(K)]E^j
=[\Delta^{-1}(K)]^{ij}E^j(K)=i\omega J^i_{ext}(K)\label{disp}.
\eea
The above expression can also be written in terms of chromodielectric tensor $\epsilon^{ij}(K)$ as
\be
[k^2\delta^{ij}-k^ik^j-\omega^2\epsilon^{ij}({\bf k},\omega)]E^j=i\omega J^i_{ext}({\bf k},\omega).
\ee
The poles of the propagator $\Delta^{ij}(K)$ give the dispersion relation 
for the waves in the medium.
When an external fast parton passes through QGP, it will disturb 
the plasma  and create induced color charge density\cite{ichimaru}. 
The total color charge density is given as 
\be
\rho^a_{tot}({\bf k},\omega)=\rho^a_{ext}({\bf k},\omega)+\rho^a_{ind}({\bf k},\omega),\label{rhot}
\ee 
where $a$ represents the color index and $\rho^{a}_{ext}({\bf k},\omega)$ is the external color charge density.
The total color charge density is linearly related to $\rho^a_{ext}$
through dielectric response function $\epsilon({\bf k},\omega)$, 
so that we may write 
\be
\rho^a_{tot}({\bf k},\omega)=\frac{\rho^a_{ext}({\bf k},\omega)}{\epsilon({\bf k},\omega)}.\label{rhoe}
\ee 
It is clearly seen that the dielectric function provides a direct 
measure of the screening of external charge density due to the induced 
color charge density in the plasma~\cite{ichimaru}.
The dielectric function can be calculated from the dielectric tensor 
using the following relation
\be
\epsilon({\bf k},\omega)=\frac{k_i\epsilon^{ij}({\bf k},\omega)k_j}{k^2}.\label{dielectric}
\ee
In an isotropic and homogeneous medium, dielectric response function 
is related to the longitudinal dielectric tensor which is independent 
of the direction of ${\bf k}$ i.e. $\epsilon({\bf k},\omega)=\epsilon_L({\bf k},\omega)$. In the general case, 
the induced color charge density is explicitly written as 
\be
\rho^a_{ind}({\bf k},\omega)=-\{1-\frac{1}{\epsilon({\bf k},\omega)}\}
\rho^a_{ext}({\bf k},\omega).\label{rind}
\ee

The wake potential induced by the fast parton is determined from the Poisson equation:
\be
\Phi^a({\bf k},\omega)=\frac{\rho^a_{ext}({\bf k},\omega)}{k^2\epsilon({\bf k},\omega)}\label{phi}.
\ee

Also one can relate the external current to the induced 
current in the linear response theory by
\be
J^i_{ind}({\bf k},\omega)=-\Pi^{ij}({\bf k},\omega)\Delta^{jk}({\bf k},\omega)J^k_{ext}.
\ee
Now we consider a charge particle, $Q^a$ moving with a constant 
velocity ${\bf v}$ and interacting with the anisotropic plasma.
The external current and charge density associated due to the test 
charge particle can be written as~\cite{plb618,prd74}:
\bea
{\bf J}^a_{ext}&=&2\pi Q^a{\bf v}\delta(\omega-{\bf k.v}),\nn\\
\rho^a_{ext}&=&2\pi Q^a\delta(\omega-{\bf k.v}).\label{ext}
\eea  
The delta function indicates that the value of $\omega$ is 
real and the velocity of the charge particle is restricted 
between $0<v<1$ which is also known as the Cerenkov condition
for the moving parton in a medium. Therefore the plasmon  
modes are determined in  the spacelike region of the 
$\omega-{\rm k}$ plane. 

\section{Self-energy in anisotropic plasma}
In this section, we shall briefly discuss how to calculate the dielectric function $\epsilon({\bf k},\omega)$ 
in an anisotropic media. The hard-loop gluon polarization tensor within  the  Valsov approximation
is given by\cite{prd68,prd62}  
\be
\Pi^{ij}(K)=-g^2\int \frac{d^3p}{(2\pi)^3}v^i\partial^lf({\bf p})
\Big(\delta^{jl}+\frac{v^jk^l}{K.V+i\epsilon}\Big),\label{self}
\ee 
where $f({\bf p})$ is the distribution function which is 
completely arbitrary. We assume that the phase-space distribution
for the anisotropic plasma is given by the following ansatz~\cite{prd68,prd70}:
\be
f({\bf p})=f_{\xi}({\bf p})=N(\xi)f_{iso}(\sqrt{{\bf p}^2+\xi({\bf p.\hat{n}})^2}).\label{dist}
\ee
$N(\xi)$ is a normalization constant, $\bf {\hat n}$ the  is direction 
of anisotropy which is along the beam direction,
$\xi$ is a parameter which represents the strength of 
anisotropy and $f_{iso}$ is an arbitrary isotropic distribution function.
Using the above ansatz one can simplify Eq.(\ref{self}) to
\be
\Pi^{ij}(K)=m_D^2\sqrt{1+\xi}\int \frac{d\Omega}{(4\pi)}
v^i\frac{v^l+\xi({\bf v.\hat{n}})n^l}{(1+\xi({\bf v.\hat{n}})^2)^2}
\Big(\delta^{jl}+\frac{v^jk^l}{K.V+i\epsilon}\Big)
\ee
where $m_D$ is the Debye mass, represented by
\be
m_D^2=-\frac{g^2}{2\pi^2}\int^{\infty}_0dpp^2\frac{df_{iso}(p^2)}{dp}
\ee
Because of the anisotropy direction, we therefore, need to construct a tensorial 
basis to represent the self-energy which depends  on not only the  
momentum $k^i$ but also the anisotropy vector $n^i$, with $n^2=1$.
Using the proper tensor basis\cite{prd68} one can decompose the 
self-energy into four structure functions as:
\be
\Pi^{ij}(k)=\alpha A^{ij}+\beta B^{ij}+\gamma C^{ij}+\delta D^{ij}\label{polarization}
\ee
$\alpha,~\beta,~\gamma$ and $\delta$ are determined by the following contractions:
\bea
k^i\Pi^{ij}k^j&=&{\bf k}^2\beta,~~~~~
\tilde{n}^i\Pi^{ij}k^j=\tilde{n}^2{\bf k}^2\delta,\nonumber\\
\tilde{n}^i\Pi^{ij}\tilde{n}^j&=&\tilde{n}^2(\alpha+\gamma),
{\rm Tr}\Pi^{ij}=2\alpha+\beta+\gamma.
\eea  
The structure functions depend on $\omega$, ${\bf k}$, $\xi$, and the  angle between the  
anisotropy vector and the momentum($\theta$). In isotropic limit($\xi\rightarrow0$) the 
structure functions $\alpha$ and $\beta$ are directly related to the isotropic 
transverse and longitudinal self-energies respectively and other structure functions
vanish~\cite{prd68}. To get the analytical expression one can calculate the  
structure functions in the small $\xi$ limit. To linear order in $\xi$ we have~\cite{prd68}
\bea
\alpha&=&\Pi_T(z)+\xi\Big[\frac{z^2}{12}(3+5\cos 2\theta)m_D^2-
\frac{1}{6}(1+\cos 2\theta)m_D^2+\frac{1}{4}\Pi_T(z)
((1+3\cos 2\theta)-z^2(3+5\cos 2\theta))\Big],\nn\\
\beta&=&z^2\Bigg[\Pi_L(z)+\xi\Big[\frac{1}{6}(1+3\cos 2\theta)m_D^2+
\Pi_L(z)(\cos 2\theta-\frac{z^2}{2}(1+3\cos 2\theta))\Big]\Bigg],\nn\\
\gamma&=&\frac{\xi}{3}(3\Pi_T(z)-m_D^2)(z^2-1)\sin^2\theta,\nn\\
\delta&=&\frac{\xi}{3k}\big[4z^2m_D^2+3\Pi_T(z)(1-4z^2)\big]\cos \theta,
\eea
with
\bea
\Pi_T(K)&=&\frac{m_D^2}{2}z^2\Big[1-\frac{1}{2}(z-
\frac{1}{z})\Big(\ln|\frac{z+1}{z-1}|-i\pi\Theta(1-z^2)\Big)\Big],\nn\\
\Pi_L(K)&=&m_D^2\Big[\frac{z}{2}\Big(\ln|\frac{z+1}{z-1}|-i\pi\Theta(1-z^2)\Big)-1\Big],
\eea
where $z=\frac{\omega}{k}$.

The dielectric tensor and the self-energy are related by the following relation:
\be
\epsilon^{ij}=\delta^{ij}-\frac{\Pi^{ij}}{\omega^2}\label{di-self}
\ee
Using Eqs.(\ref{dielectric}) and (\ref{polarization}) together with Eq.(\ref{di-self}), one can find 
the dielectric function in terms of structure functions in AQGP.
\section{Induced charge density}
In the presence of the test charge particle, the induced charge density 
and the wake potential depend on the velocity of the test charge and
the distribution of the background particle~\cite{ppp}. When a charge 
particle is introduced into a plasma at rest, it acquires a shielding cloud. 
As a result, the induced charge distribution is spherically symmetric.
When a charge particle is moving with a fixed velocity in the plasma, the 
induced charge distribution no longer remains symmetric. It deforms the screening 
cloud from a spherical shape to an ellipsoidal one. Therefore the induce 
charge distribution loses forward and backward symmetry. 

We assume that in the AQGP, the average phase velocity is $v_p$. 
According to Cerenkov condition there will be two important scenarios 
which occur due to the interaction of the particle with the plasmon wave: 
First, the modes with low speed(less than the plasmon phase velocity) can
be excited, but the particles moving slightly slower than the wave, will be 
accelerated while the charge particle moving faster than the wave 
will decrease its average velocity~\cite{ichimaru}. The slowly moving particle 
absorbs energy from the wave and the faster particle transfers its extra energy
to the wave. The absorption and emission of energy result in a wake
in the induced charge density as well as in the potential. Second, when the charge 
particle moving with a speed greater than the average phase velocity $v_p$,
the modes are excited and they may not be damped. The excited modes can 
generate Cerenkov-like radiation and Mach stem which leads to oscillation
both in induced charge density and in wake the potential. It is well known~\cite{Landau}
that a parton moving with a supersonic speed in a plasma produces a Mach 
region which is of conical shape with an opening angle with respect to the 
direction of the particle propagation given by the expression

\begin{figure}[t]
\begin{center}
\epsfig{file=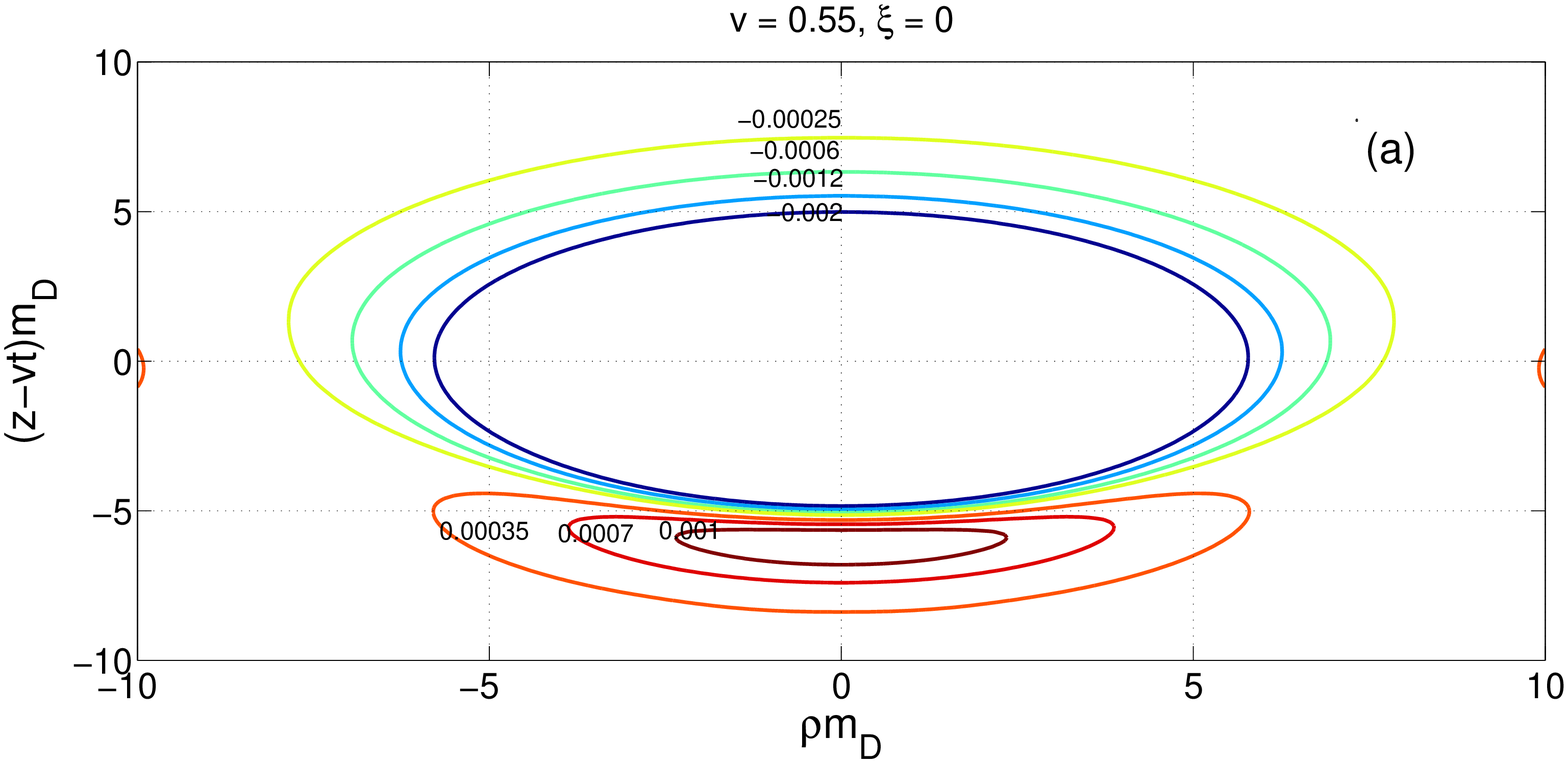,width=6.5cm,height=6.5cm,angle=0}~\epsfig{file=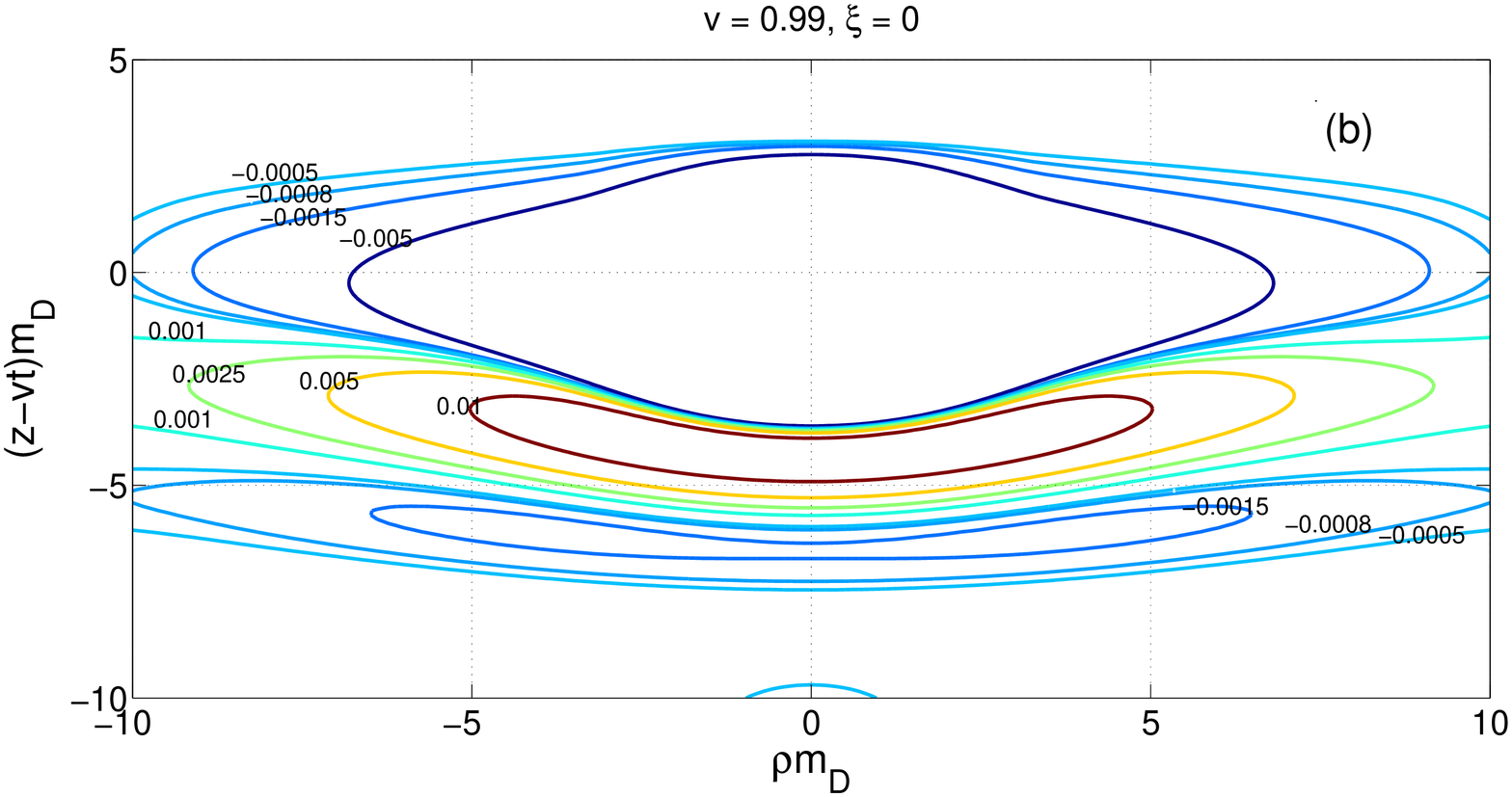,width=6.5cm,height=6.5cm,angle=0}
\epsfig{file=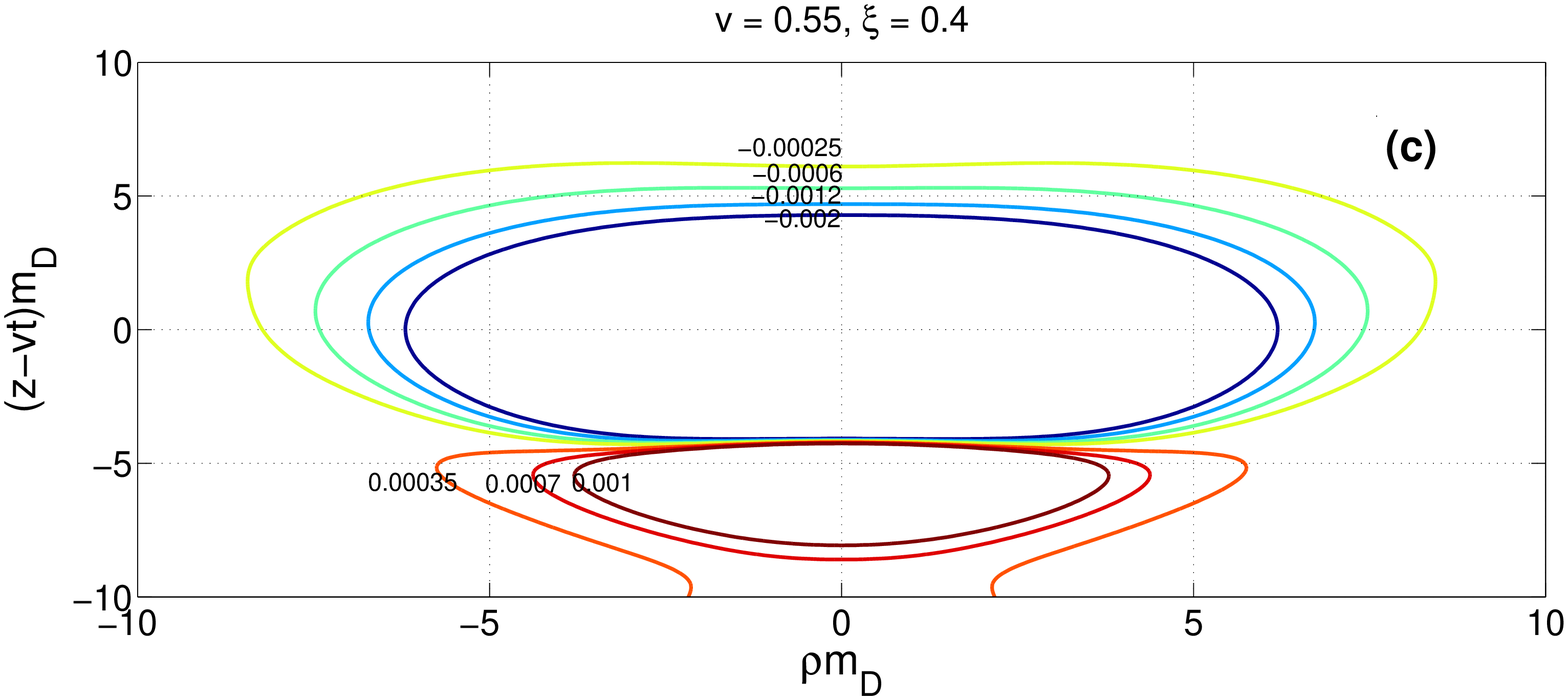,width=6.5cm,height=6.5cm,angle=0}~\epsfig{file=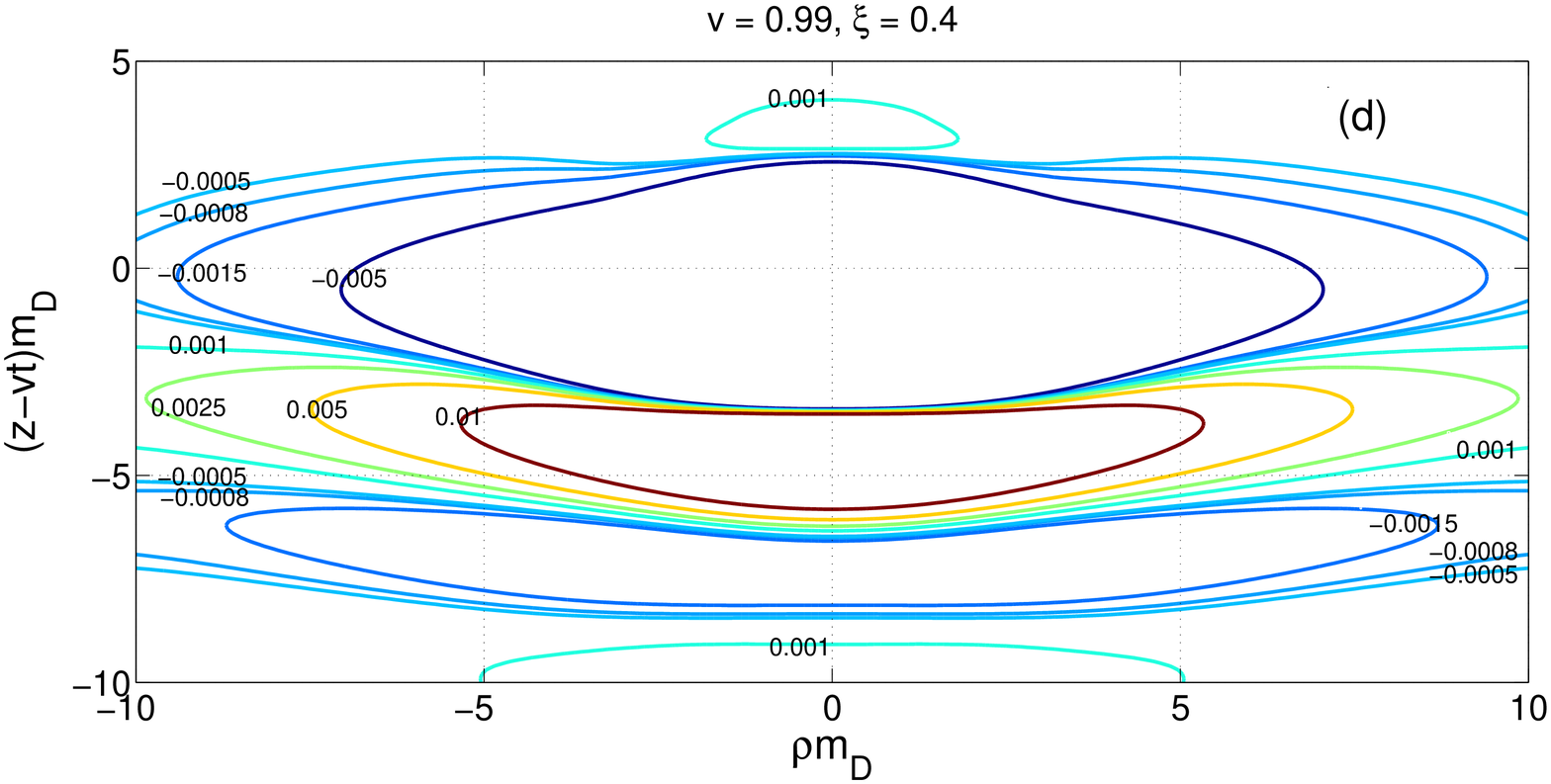,width=6.5cm,height=6.5cm,angle=0}
\epsfig{file=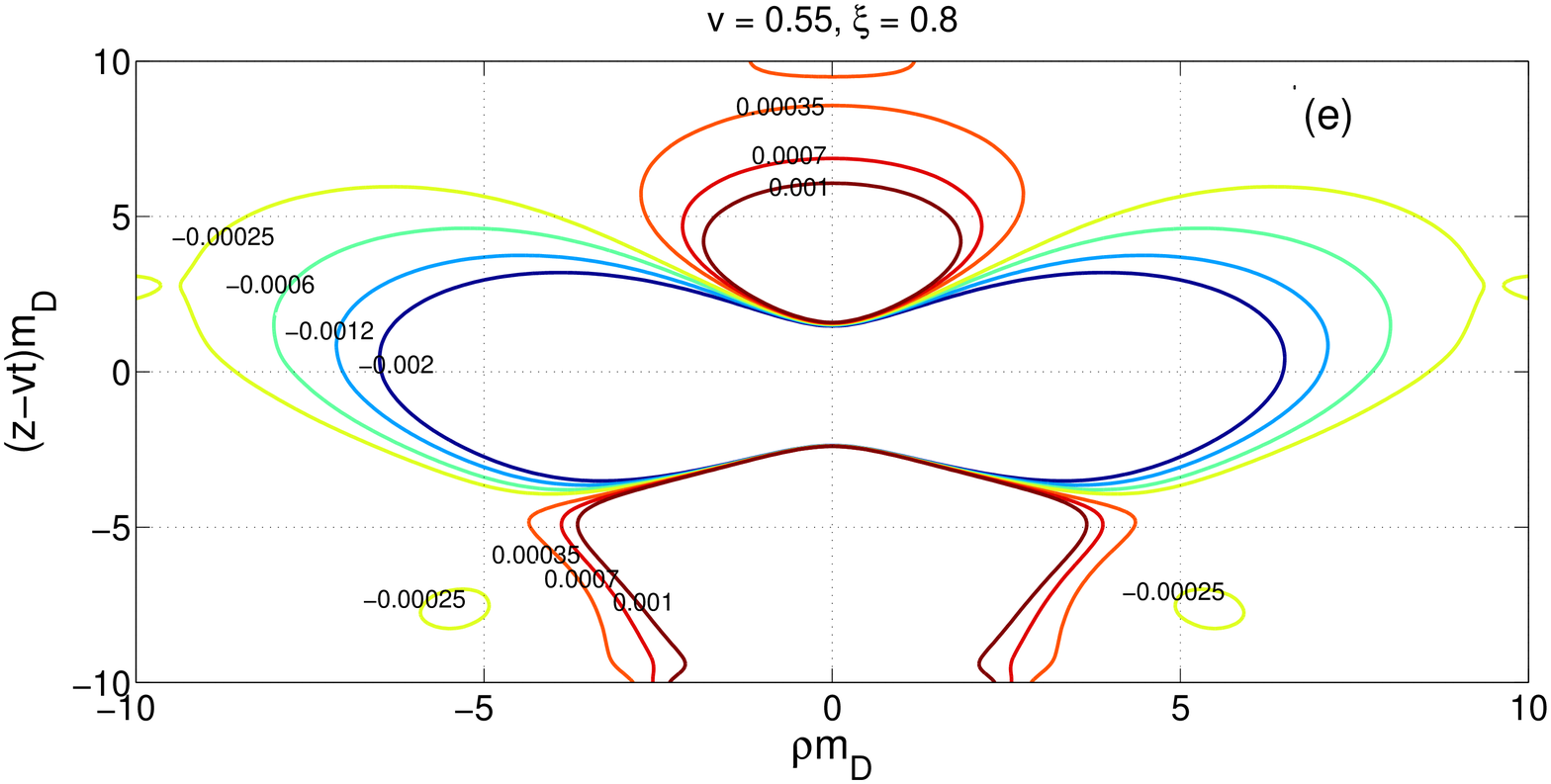,width=6.5cm,height=6.5cm,angle=0}~\epsfig{file=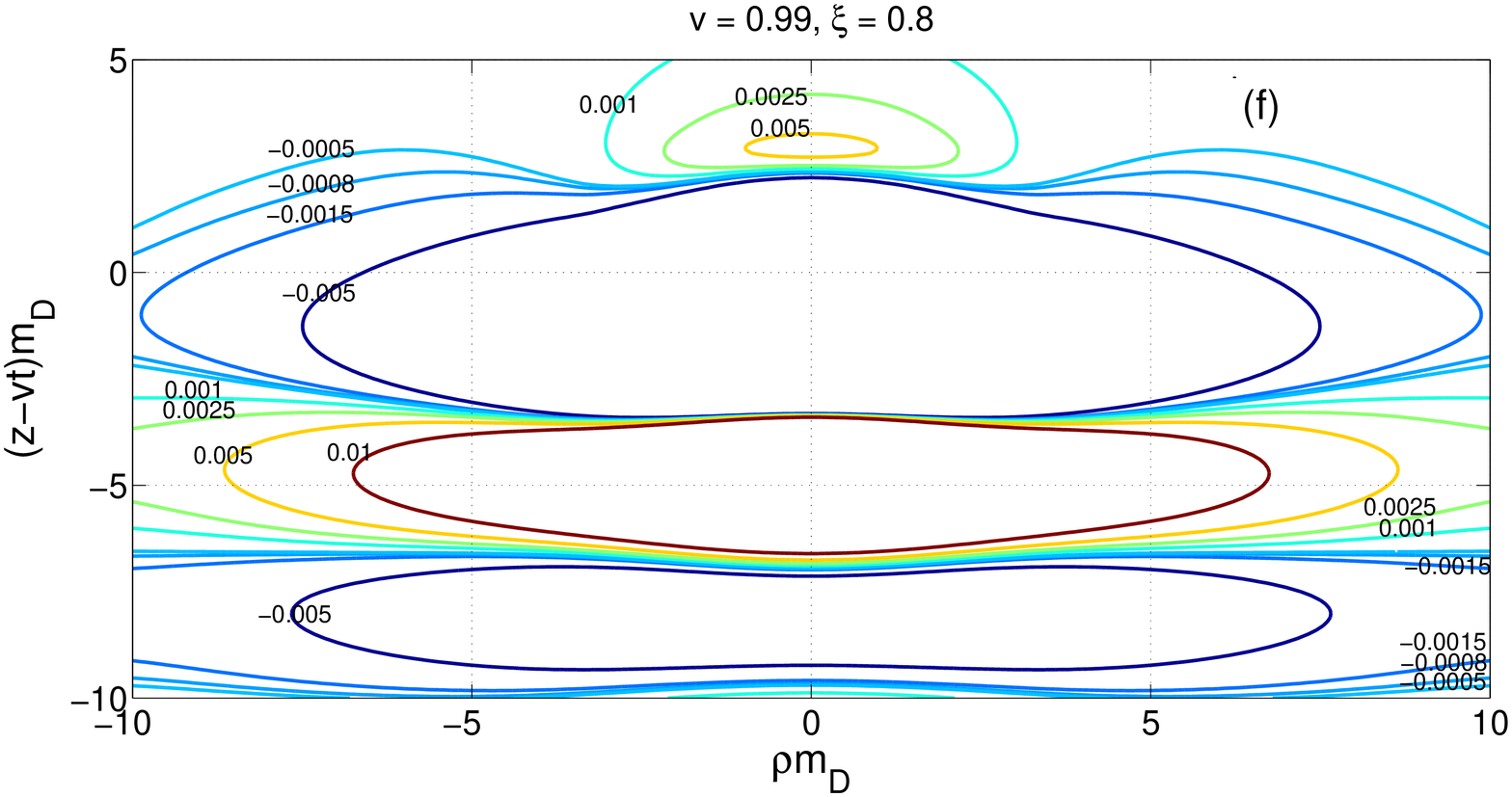,width=6.5cm,height=6.5cm,angle=0}
\end{center}
\caption{(Color online) Laft Panel: The plot shows equicharge line with  parton velocity 
$v=0.55$ for different $\xi (0, 0.4, 0.8)$. Right Panel: Same as left panel with parton
velocity $v=0.99$.}  
\label{fig1}
\end{figure}

\begin{figure}[t]
\begin{center}
\epsfig{file=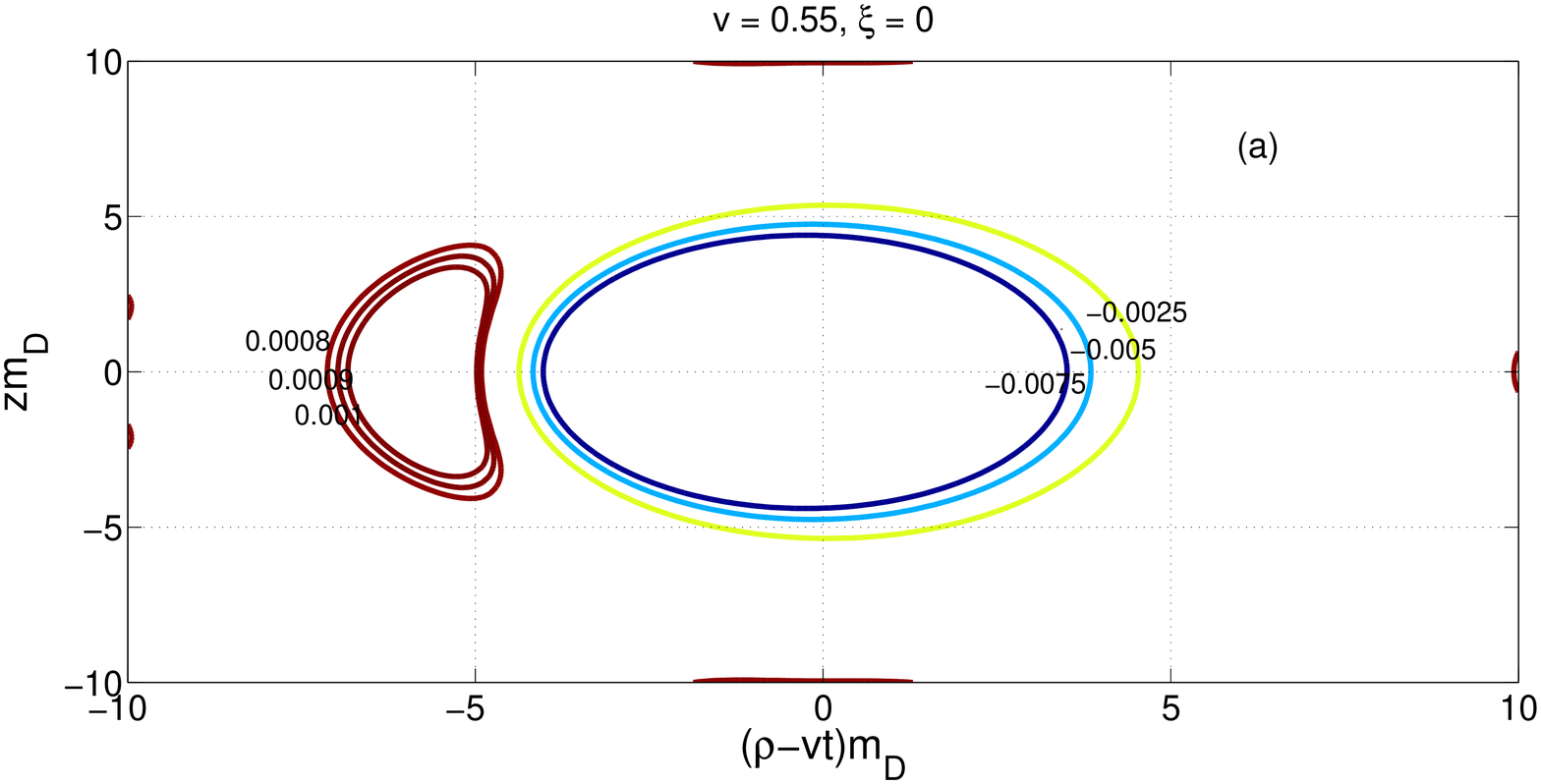,width=6.5cm,height=6.5cm,angle=0}~\epsfig{file=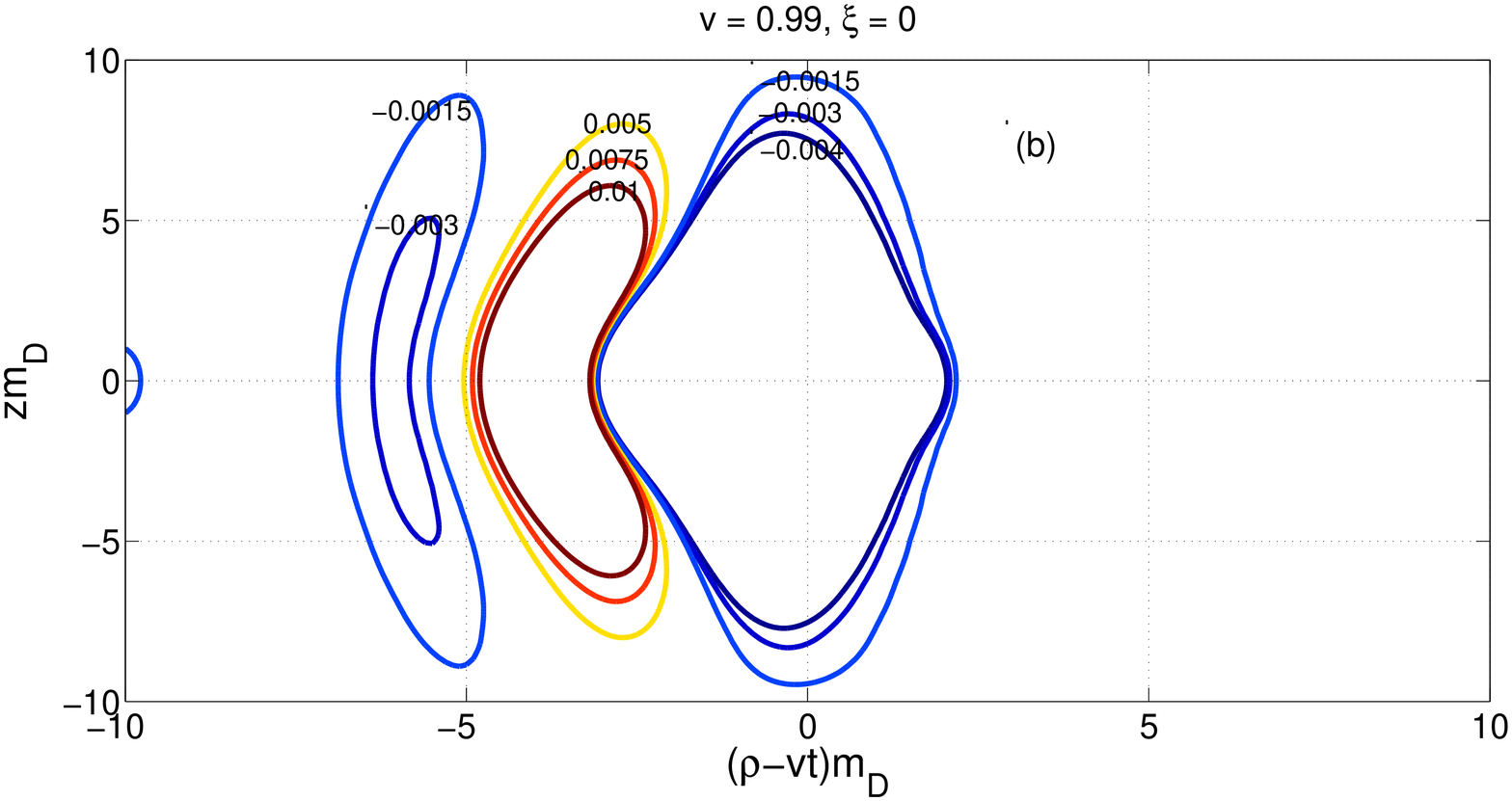,width=6.5cm,height=6.5cm,angle=0}
\epsfig{file=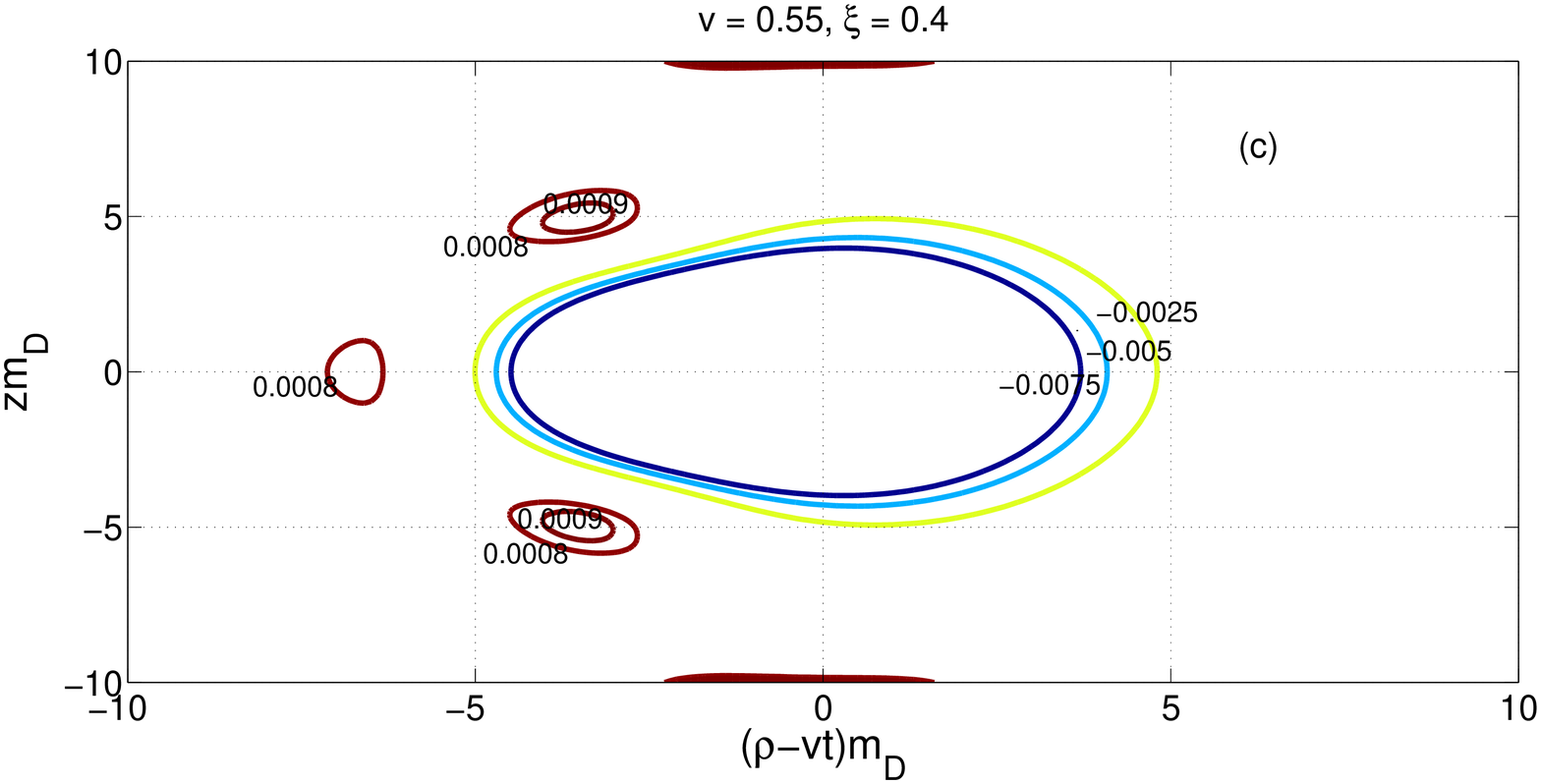,width=6.5cm,height=6.5cm,angle=0}~\epsfig{file=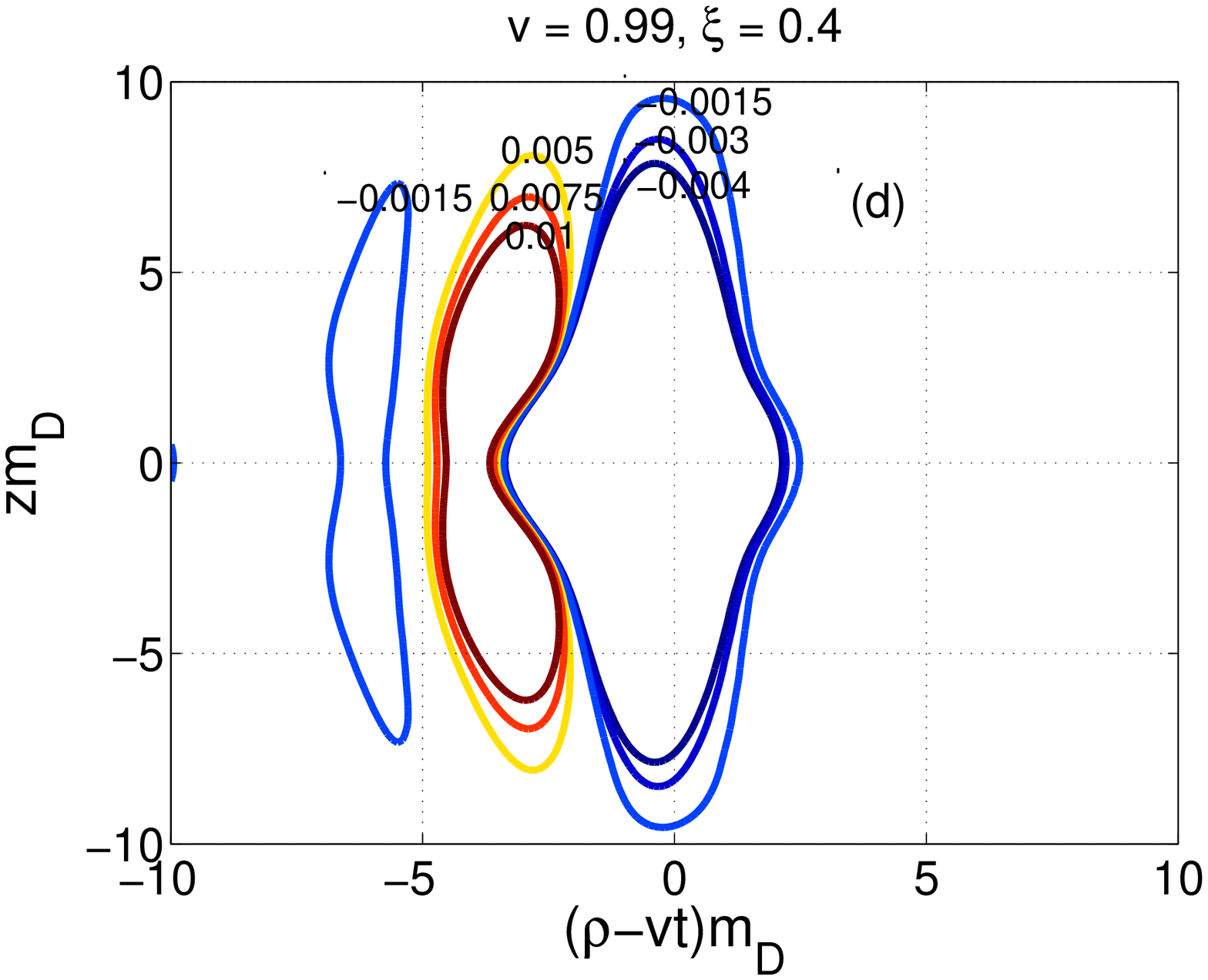,width=6.5cm,height=6.5cm,angle=0}
\end{center}
\caption{(Color online) The left(right) panel is shows the equicharge lines for 
$v = 0.55(0.99)$. In this case the parton moves perpendicular to the direction of anisotropy.
}  
\label{fig_new}
\end{figure}

\be
\theta_M=\arcsin(\frac{c_s}{v}),
\ee
where $c_s$ is the sound velocity. The cone of the front of the shock wave has the angle 
$\theta_f=\pi/2-\theta_M$~\cite{npa767}. Therefore, the cone of the particles should be 
produced, when the parton is  moving with a supersonic speed in a plasma.

Now we study the color charge density induced by the fast parton in an anisotropy media.
By substituting Eq.(\ref{ext}) into Eq.(\ref{rind})  and transforming into ${\bf r}-t$ space,
the induced charge density becomes,
\bea
\rho^a_{ind}({\bf r},t)=2\pi Q^a\int\frac{d^3k}{(2\pi)^3}\int\frac{d\omega}
{2\pi}{\rm exp}^{i({\bf k.r}-\omega t)}
(\frac{1}{\epsilon({\bf k},\omega)}-1)\delta(\omega-{\bf k.v}).
\eea
We assume that the parton is moving along the z-direction which is the beam direction i.e. ${\bf v}||{\bf \hat n}$. 
We use spherical coordinates system for ${\bf k}$ i.e. ${\bf k}=(k\sin\theta\cos\phi,k\sin\theta\sin\phi,k\cos\theta)$ 
and cylindrical coordinates for ${\bf r}=(\rho,0,z)$. The induced charge density can be written as
\bea
\rho^a_{ind}({\bf r},t)=\frac{Q^am_D^3}{2\pi^2}\int^\infty_0 dk~k^2 
\int^{1}_{0}d\chi~J_0(k \rho\sqrt{1-\chi^2}m_D)
\Big[\cos\Gamma(\frac{{\rm Re}\epsilon({\bf k},\omega)}
{\Delta}-1)+\sin\Gamma\frac{{\rm Im}\epsilon({\bf k},\omega)}{\Delta}\Big]\Bigg{|}_{\omega={\bf k.v}},\label{ind_charge}
\eea
where $\chi$ is represented as $\cos\theta$, $J_0$ is the zeroth-order 
Bessel function, $\Gamma=k\chi(z-vt)m_D$ and 
$\Delta=({\rm Re}\epsilon({\bf k},\omega))^2+({\rm Im}\epsilon({\bf k},\omega))^2$.
To get the above equation we use the simple transformation $\omega\rightarrow \omega m_D$ and 
$k\rightarrow km_D$. It is seen that the charge density $\rho^a_{ind}$ is
proportional to $m_D^3$. We now discuss the numerical result of the induced color charge 
density with two different speeds of the fast parton: One is below the average phase velocity($v=0.55$)
and other is greater than it($v=0.99$). 

In Figs.~(1a-1f) we display the contour plot of the scaled equicharge
lines for isotropic and anisotropic plasma.
The contour plot of the equicharge lines shows a sign flip along the direction of the 
moving parton in Fig.\ref{fig1}.
The left(right) panel in Fig.\ref{fig1} shows the contour plot of the induced color charge density
in isotropic and anisotropic plasma for two different anisotropic parameter $\xi$ with parton velocity 
$v=0.55(0.99)$. It is clearly seen that the equicharge lines are modified in anisotropic plasma for 
parton velocity $v=0.55$. With the increase of strength of the anisotropy we found that the positive charge 
lines appear alternately in the backward space which leads to small oscillatory behavior of the 
color charge wake(see in Fig.(1e)). 
 
When a charge particle moves faster than the average speed of the plasmon,
the induced charge density forms a cone like structure as shown in the right panel of Fig.\ref{fig1}.
It is clearly noticed that the color charge wake is significantly different from when the parton velocity is $v=0.55$.
It is also seen that the induced charge density is oscillatory in nature. The supersonic nature 
of the parton leads to formation of the Mach cone and the plasmon modes could emit a Cerenkov-like 
radiation, which spatially limits the disturbances in the induced charge density ~\cite{ppcf}.
Due to the effect of the anisotropy, the color charge wake is modified significantly and the oscillatory 
behavior is more pronounced than the isotropic case and  it is also seen that the oscillatory nature
increases with the increase of the  anisotropic parameter $\xi$. In the backward space $((z-vt)<0)$, induced color 
charge density is very much more sensitive to the anisotropic plasma than that in the forward space$((z-vt)>0)$. 
If we consider charge particle moves in a plane, the structure of the contour plot is the same as Fig.\ref{fig1}; 
the only difference is that  its makes an angle within the  $\rho-z$ plane.

Next we consider the case when the parton moves perpendicular
to the anisotropy direction in which case  
the induced charge density can be written as
\bea
\rho^a_{ind}({\bf r},t)=\frac{Q^am_D^3}{2\pi^2}\int^\infty_0 dk~k^2
\int^1_0d\chi\int^{2\pi}_0\frac{d\phi}{2\pi}\Big[\cos\Omega(\frac{{\rm Re}\epsilon({\bf k},\omega)}{\Delta}-1)
+\sin\Omega\frac{{\rm Im}\epsilon({\bf k},\omega)}{\Delta}\Big]\Bigg{|}_{\omega={\bf k.v}},
\eea
with $\Omega = k(z\chi +(\rho-vt)\sqrt{1-\chi^2}\cos\phi)m_D$.
Numerical evaluation of the above equation leads to the contour plots of the induced charge density shown in 
Fig.~(2a-2d). The left(right) panels show the contour plots of the induced charge density for the parton 
velocity is $v = 0.55(0.99)$. Because of the effect of anisotropy, one observes 
a clear modification of the 
induced charge density. When $v = 0.99$, i.e. $v$ is larger than the phase 
velocity $v_p$, the number of 
induced charged lines that appear alternately in the backward space is 
reduced for the anisotropic plasma in comparison to the
isotropic plasma. 
Therefore the anisotropy reduces the oscillatory behavior of the induced color 
charge density, when the parton moves in the transverse plane.
This, as we shall
see later,  will lead to less oscillatory wake potential.
Note that these observations are just opposite to the case when the parton
moves parallel to the anisotropy direction.

\section{Wake potential in AQGP}
By combining Eqs.(\ref{phi}) and (\ref{ext}), the wake potential in $r-t$ space due to the motion 
of a charge parton can be written as~\cite{plb618,prd74}
\bea
\Phi^a({\bf r},t)=2\pi Q^a\int \frac{d^3k}{(2\pi)^3}\int 
\frac{d\omega}{2\pi}~{\rm exp}^{i({\bf k.r}-\omega t)}
\frac{1}{k^2\epsilon(\omega,{\bf k})}\delta(\omega-{\bf k.v}).
\eea
Using similar  coordinate system as before, the screening potential turns into
\bea
\Phi^a({\bf r},t)=\frac{Q^am_D}{2\pi^2}\int^{\infty}_0 dk\int^{1}_{0} d\chi J_0(k \rho\sqrt{1-\chi^2}m_D)
\Big[\cos\Gamma\frac{{\rm Re}\epsilon(\omega,{\bf k})}{\Delta}+
\sin\Gamma\frac{{\rm Im}\epsilon(\omega,{\bf k})}{\Delta}\Big]\Bigg{|}_{\omega={\bf k.v}}.\label{screening_pot}
\eea
The above equation shows that the wake potential is proportional to the Debye mass.
We solve the wake potential for two spatial cases, 
(i) along the parallel direction of the fast parton, i.e. ${\bf r}\parallel{\bf v}$ and also
$\rho=0$ and
(ii) perpendicular to direction of the parton, i.e. ${\bf r\perp v}$.
The potential for the first case is obtained as 
\bea
\Phi^a_{\parallel}({\bf r},t)=\frac{Q^am_D}{2\pi^2}\int^{\infty}_{0}dk
\int^{1}_0d\chi
\Big[\cos\Gamma\frac{{\rm Re}\epsilon(\omega,{\bf k})}{\Delta}+
\sin\Gamma\frac{{\rm Im}\epsilon(\omega,{\bf k})}{\Delta}\Big]\Bigg{|}_{\omega={\bf k.v}}, 
\eea
whereas that for the perpendicular case is
\bea
\Phi^a_{\perp}({\bf r},t)=\frac{Q^am_D}{2\pi^2}\int^{\infty}_{0}dk
\int^{1}_0d\chi J_0(k \rho\sqrt{1-\chi^2}m_D) 
\Big[\cos\Gamma^{\prime}\frac{{\rm Re}\epsilon(\omega,{\bf k})}{\Delta}-
\sin\Gamma^{\prime}\frac{{\rm Im}\epsilon(\omega,{\bf k})}{\Delta}\Big]\Bigg{|}_{\omega={\bf k.v}},
\eea
with $\Gamma^{\prime}=k\chi vtm_D$.

\begin{figure}[t]
\begin{center}
\epsfig{file=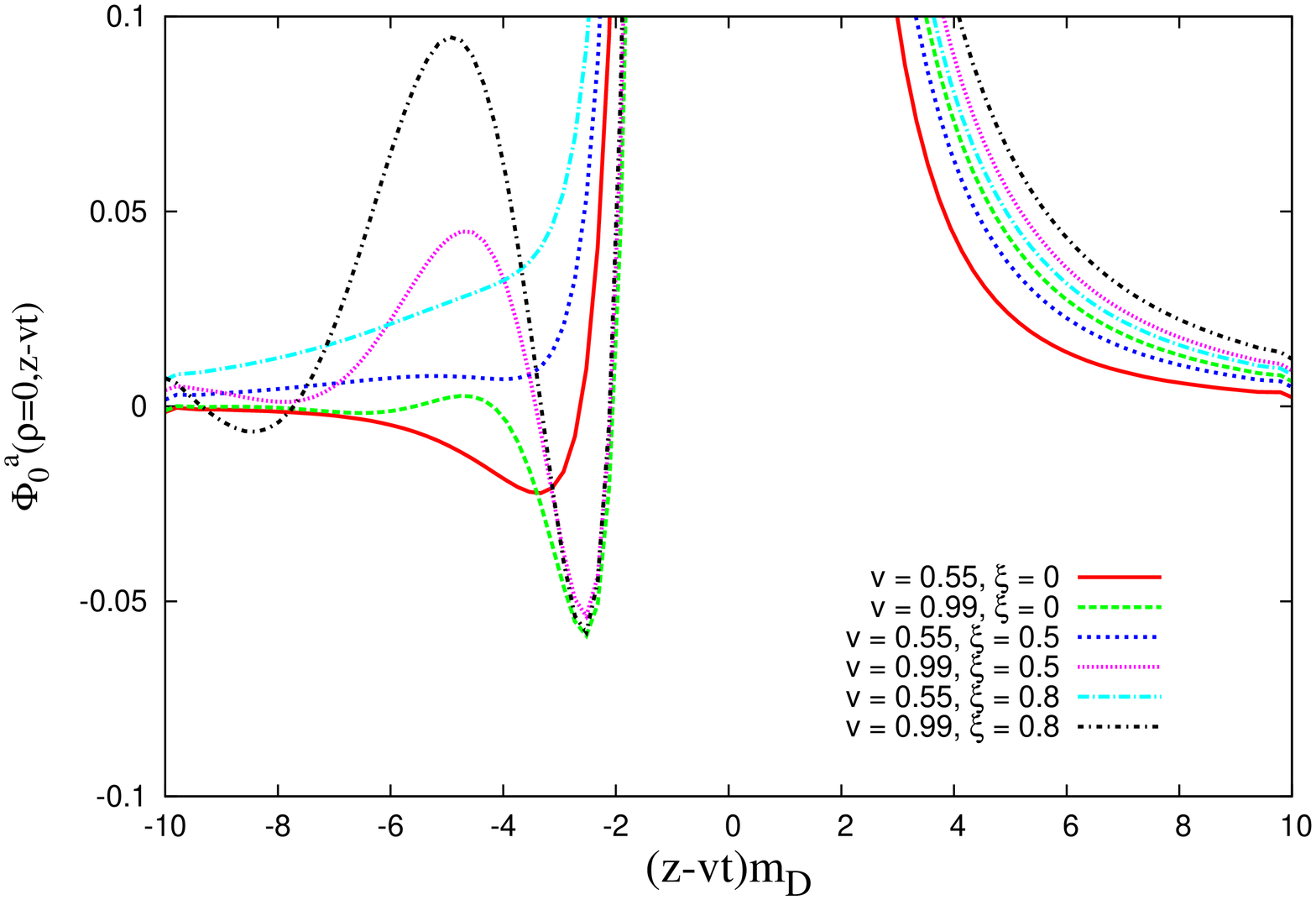,width=7.5cm,height=6.5cm,angle=270}~\epsfig{file=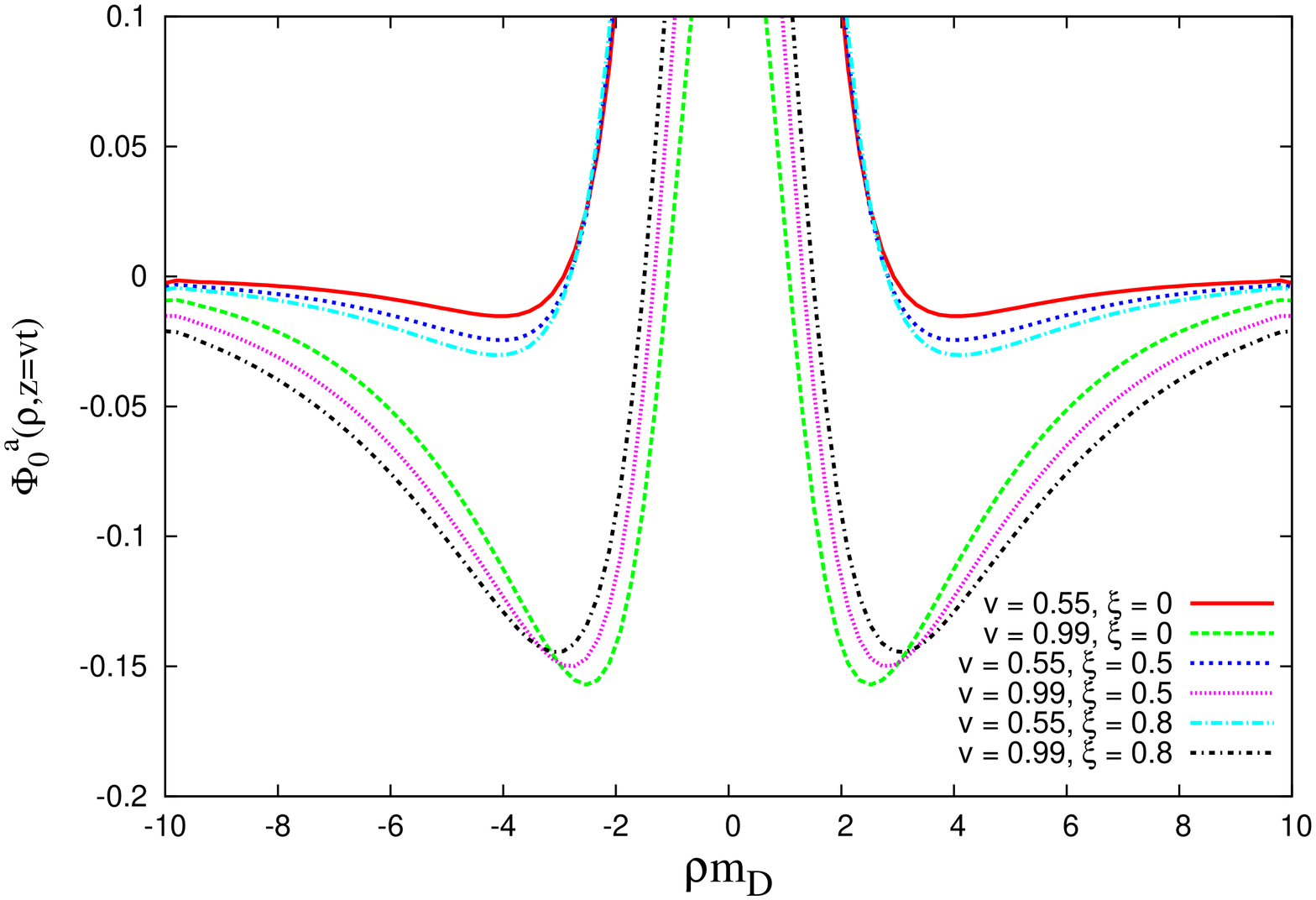,
width=7.5cm,height=6.5cm,angle=270}
\end{center}
\caption{(Color online) Left panel: Scaled wake potential along the motion of the fast parton i.e. $z-$axis 
for different $\xi$ with two different parton velocity $v=0.55$ and $v=0.99$. Right panel: same as left panel but 
perpendicular to direction of motion of the parton.
}
\label{fig2}
\end{figure}

Fig. \ref{fig2} describes the wake potential in two specific directions. In these figures, the scaled parameter
$\Phi^a_0$ is given by $\frac{2\pi^2}{m_D}\Phi^a$. The left panel shows the wake potential along the direction 
of the moving color charge. In the backward space, the wake potential for isotropic plasma decreases with the 
increase of $z-vt$ and exhibits a negative minimum when $v=0.55$ i.e. in the backward direction, the wake 
potential is a Lennard-Jones potential type which has a short range repulsive part as well as a long range 
attractive part~\cite{prd74}. With the increase of the anisotropic parameter $\xi$ the depth of the negative 
minimum decreases for $v=0.55$. The position of the negative minimum is same for $\xi=0$ and $\xi=0.5$. 
But for $\xi=0.8$, the wake potential increases towards the origin, and their are no minima and behaves like a 
modified Coulomb potential. When the parton moves faster than $v_p$, i.e. when $v=0.99$, the wake potential is 
much more different than in the case of $v=0.55$. For $v=0.99$, the wake potential shows an oscillatory behavior 
in both isotropic and anisotropic plasma. Such oscillation of the wake potential is clearly reflected only in 
the backward direction. It is clearly visible that the depth of negative minimum is increased and it is shifted 
towards the origin compared to the case when $v=0.55$. Due to the anisotropic effect of the plasma, the 
oscillation of the wake potential is more pronounced and it extends to a large distance. But, there is no 
significant change of the absolute value of the negative minimum for different $\xi(0.5,0.8)$. In the forward 
direction, the screening potential is a modified Coulomb potential. Moreover, with the increase of $\xi$ the 
potential increases for both $v=0.55$ and $v=0.99$. The right panel of Fig. \ref{fig2} describes the wake 
potential along the perpendicular direction of the moving parton. It can be seen  that the wake potential 
is symmetric in backward and forward directions, no matter what the speed is. The structure of the 
wake potential is found to be the Lennard-Jones type due to the deformed screening charge cloud  in the presence of
the moving charge particle. When $v=0.55$, the value of negative minimum is increased with an increase of $\xi$, but 
in the case of $v=0.99$, it decreases with $\xi$. However, with the increase of $\xi$ the depth of negative minimum
is moving away from the origin for both the jet velocities considered here.

\begin{figure}[t]
\begin{center}
\epsfig{file=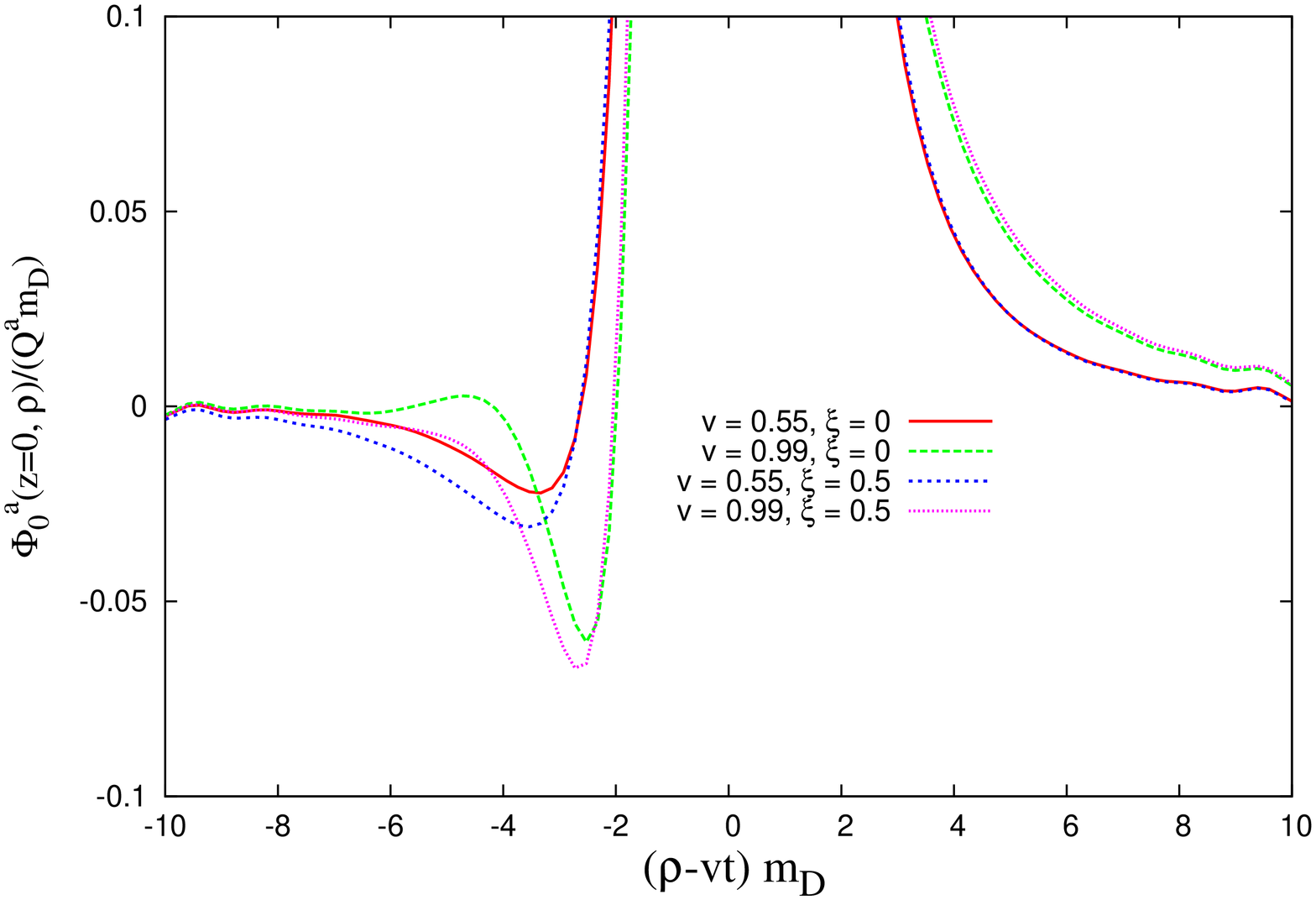,width=7.5cm,height=6.5cm,angle=270}~\epsfig{file=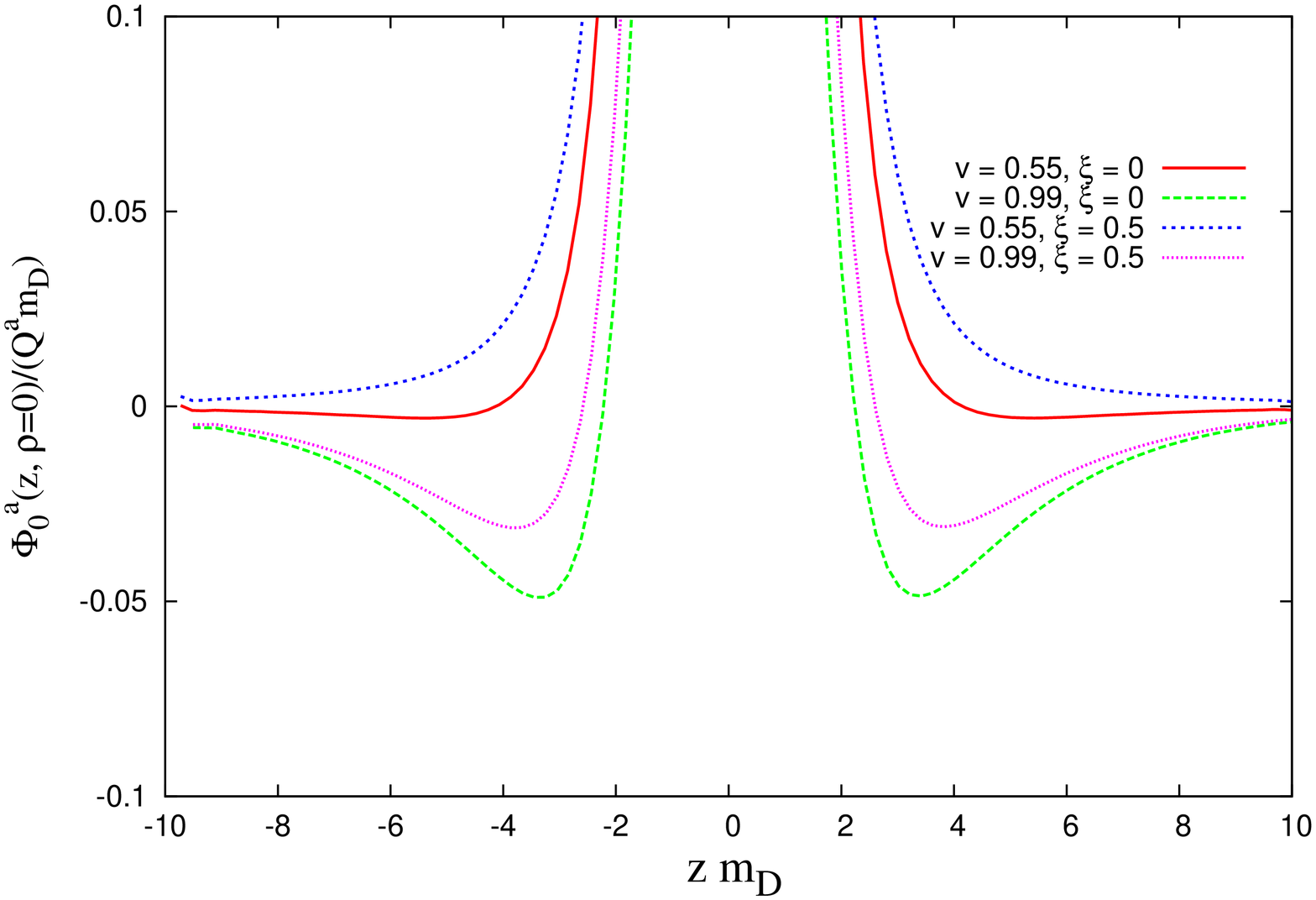,
width=7.5cm,height=6.5cm,angle=270}
\end{center}
\caption{(Color online) The left(right) panel shows scaled wake potential  
for $\xi = \{0, 0.5\}$ with parton velocity $v=0.55(0.99)$. In this case 
the parton moves perpendicular to the direction of anisotropy.
}
\label{fig3}
\end{figure}

We now discuss the wake potential when the parton moves perpendicular to the
anisotropy direction. 
The wake potential in (\ref{screening_pot}) is also solved for two 
spatial cases: (i) along the direction
of the moving parton i.e. ${\bf r}\parallel{\bf v}$ and (ii) perpendicular direction of the parton, 
i.e. ${\bf r\perp v}$. The wake potential for the parallel case can be
 written as 
\bea
\Phi^a_{\parallel}({\bf r},t)=\frac{Q^am_D}{2\pi^2}\int^{\infty}_{0}dk
\int^{1}_0d\chi\int^{2\pi}_0\frac{d\phi}{2\pi}
\Big[\cos\Omega^{\prime}\frac{{Re}\epsilon(\omega,{\bf k})}{\Delta}+
\sin\Omega^{\prime}\frac{{\rm Im}\epsilon(\omega,{\bf k})}{\Delta}\Big]\Bigg{|}_{\omega={\bf k.v}},
\eea
where $\Omega^{\prime} = k(\rho-vt)\sqrt{1-\chi^2}\cos\phi m_D$.
For the perpendicular case it is given by,
\bea
\Phi^a_{\perp}({\bf r},t)=\frac{Q^am_D}{2\pi^2}\int^{\infty}_{0}dk
\int^{1}_0d\chi\int^{2\pi}_0\frac{d\phi}{2\pi}
\Big[\cos\Omega^{\prime\prime}\frac{{Re}\epsilon(\omega,{\bf k})}{\Delta}+
\sin\Omega^{\prime\prime}\frac{{\rm Im}\epsilon(\omega,{\bf k})}{\Delta}\Big]\Bigg{|}_{\omega={\bf k.v}},
\eea
with $\Omega^{\prime\prime} = k(z\chi-vt\sqrt{1-\chi^2}\cos\phi)m_D$.
In Fig. \ref{fig3}, the scaled wake potentials are shown for both parallel 
and  perpendicular directions of 
the parton i. e. when ${\bf v}\parallel {\bf r}$ and ${\bf v}\perp {\bf r}$. 
The left panel shows  
screening potential along the parallel direction of the moving color charge.
For both the velocity limits considered her, in the forward (backward) direction, 
the behavior of the wake potential is more 
like a modified Coulomb (Lennard-Jones) potential as in the case when 
${\bf v}\parallel {\bf \hat n}$.
For $v = 0.99$, the wake potential shows an oscillatory behavior in an isotropic plasma~\cite{prd74} but in 
anisotropic case, oscillatory structure of the wake potential is smeared out 
for $\xi = 0.5$ and ${\bf v}\perp {\bf \hat n}$. 
It is clearly seen that the depth
of the negative minimum is increased in the case of anisotropic plasma 
for both the parton velocities considered here.
Moreover, in the anisotropic case the depth of the negative minimum is moving away from the origin.
The contribution
of the wake potential in the perpendicular direction of the moving parton 
is shown in the right panel in 
Fig. \ref{fig3}. When the parton moves with the velocity $v = 0.55$, 
the anisotropy modifies the structure of the 
wake potential significantly; i.e. it becomes a modified coulomb potential
instead of Lennard-Jones potential. 
For $v = 0.99$, the wake potential is a
Lennard-Jones potential type but the depth of the minimum decreases 
in anisotropic plasma.

\section{Summary} 
In this work, we have investigated the wake in charge density as well as
the wake potential induced by a fast 
parton propagating through the anisotropic quark-gluon plasma expected to be 
formed in heavy-ion collisions. For the sake of simplicity small $\xi$ expansion
($0<\xi < 1$)
of the polarization tensor has been considered. The anisotropic 
effect modifies the dielectric tensor of the QGP, as a consequence, it 
influences the distribution of the induced
color charge and the screening potential. 
To calculate the induced charge density of AQGP, two different velocities 
of the parton have been used for both  
the induced charge density and the wake potential. When the parton moves 
parallel to the anisotropy direction with a speed $v=0.55$, the anisotropic 
effect at $\xi=0.8$ makes a small oscillation of the induced charge density in
contrast to the isotropic case. For a larger speed $v=0.99$, which
is greater than the average plasmon speed, the effect of the anisotropy 
on the charge density becomes more remarkable. 
We also show that with the increase of $\xi$, the distribution of the 
induced charge density is much more oscillatory in nature. For the case of 
the wake potential, we focus on the parallel and perpendicular directions of 
the fast parton. Along the parallel direction of parton, the wake potential 
shows a decrease of negative minimum in presence of anisotropy in the backward 
direction and the potential at $\xi=0.8$ behaves as a modified Coulomb-like 
potential for $v=0.55$. For $v=0.99$, the wake potential exhibits an 
oscillatory behavior in the backward 
direction and it is amplified with the increase of the strength of the 
anisotropy but no appreciable change of the negative 
minimum is observed. The numerical analysis of the wake potential along 
the perpendicular direction of moving parton leads to
Lennard-Jones type potential for both  $v=0.55$ and $v=0.99$. 
Furthermore, we have investigated the case when ${\bf v}\perp{\bf \hat n}$.
Results show that the anisotropy minimizes the oscillatory strength of the induced
charged density, when the parton moves with the  velocity $v = 0.99$.
As a consequence we do not find any oscillatory nature of the 
wake potential in the backward direction at $\xi = 0.5$, when the 
parton moves along the parallel direction with $v = 0.99$. For $\xi = 0.5$, 
the depth of the negative minimum is increased and also shifts away from 
the origin for both the parton velocities considered here. 
In the perpendicular direction of the moving parton with velocity
$v = 0.55$ and $\xi=0.5$, the screening potential behaves like modified Coulomb
potential instead of a Lenard-Jones potential, contrary to
the case when ${\bf v}\parallel{\bf \hat n}$.

We end by making the following comments. The effect of collision on the
collective modes in anisotropic QGP has been investigated by using
BGK collisional kernel where the wave vector is parallel to the
anisotropy direction~\cite{prd73}. The reason for this is that the growth of the
unstable modes is the highest in such case. It is found that the inclusion
of collision slows down the growth rate of the unstable modes. Similar
effects can also be incorporated in the present calculation as has been
done in Ref.~\cite{jpg34} for the isotropic case where it has been shown that
the wakes are significantly modified. It is thus expected that inclusion
of collision in the case of AQGP might lead to further modifications, such
as, for both the velocity limits considered here, we expect more oscillatory 
behavior when $v\parallel n$ and almost non-oscillatory nature in the
wake potential in the backward direction for $v>v_p$ 
when $v\perp {\hat n}$. Work on this issue is in progress~\cite{MM1}.
It is to be noted that we have considered here the small $\xi$ case. The 
extension of the present calculation for arbitrary $\xi$ is
worth investigating.


\begin{thebibliography}{50}
\bibitem{jetquen1} J. D. Bjorken, Report No. Fermilab-Pub-82/59-THY(1982) and
Erratum (unpublished).  
\bibitem{jetquen2} M. Gyulassy, P. Levai and I. Vitev, Nucl. Phys. B
{\bf 571}, 197 (2000). 
\bibitem{jetquen3} B. G. Zakharov, JETP Lett. {\bf 73}, 49 (2001).
\bibitem{jetquen4} M. Djordjevic and U. Heinz, Phys. Rev. Lett {\bf 101},
022302 (2008). 
\bibitem{jetquen5} G -Y Qin, J. Ruppert, C. Gale, S. Jeon, G. Moore, and
M. G. Mustafa, Phys. Rev. Lett {\bf 100}, 072301 (2008).
\bibitem{jetquen6} R. Baier {\it et al.}, J. High Energy. Phys.  {\bf 09}, 033 (2001). 
\bibitem{jetquen7} S. Jeon and G. D. Moore, Phys. Rev. {\bf C 71}, 034901 (2005).  
\bibitem{jetquen8} A. K. Dutt-Mazumder, J. Alam, P. Roy, and B. Sinha,
Phys. Rev. D {\bf 71}, 094016 (2005). 
\bibitem{jetquen9} P. Roy, J. Alam, and A. K. Dutt-Mazumder,J. Phys. G. {\bf 35}, 104047 (2008).

\bibitem{prl95} J. Adams, et al., STAR collaboration, Phys, Rev. Lett {\bf 95} (2005) 152301.
\bibitem{prl97} S. S Adler, et al., PHENIX collaboration, Phys, Rev. Lett {\bf 97} (2006) 052301.

\bibitem{pg_34} J. Casalderry-Solana, J. Phys. {\bf G34},S345 (2005).

\bibitem{yy} J. Rupport, Nucl. Phys. {\bf A774}, 397 (2007).

\bibitem{prl103} J. Takahashi, B. M. Travares, W. L. Qian, R. Andrade, 
F. Grassi, Y. Hama, T. Kodama, and N. Xu, Phys. Rev. Lett {\bf 103}, 242301
(2009).

\bibitem{prc86} G. Aad {\it et al}, Phys. Rev. {\bf C86}, 014907 (2012). 

\bibitem{plb618} J.~Ruppert and B.~Muller Phys.~Lett. B {\bf 618}, 123 (2005).
\bibitem{prd74} P.~Chakraborty,~M.~G.~Mustafa and M.~H.~Thoma,~Phys. Rev. D {\bf 74}, 094002 (2006).
\bibitem{jpg34} P.~Chakraborty, et al., J.~Phys.~G {\bf 34}, 2141 (2007).
\bibitem{npa856} Bing-feng Jiang, Jia-rong Li, Nucl. Phys. A {\bf 856},  121 (2011).
\bibitem{jpg39}  Bing-feng Jiang, Jia-rong Li, J.Phys. G {\bf 39}, 025007 (2012).

\bibitem{npa750} H.~Stoecker~Nucl.~Phys.~A {\bf 750}, 121 (2005).
\bibitem{jconf} J.~Casalderrey-Solana,~E.~V.~Shuryak and D.~Teaney, J.~Conf.~Ser. {\bf 27}, 22 (2005).

\bibitem{prl96} V.~Koch,~A.~Majumder and Xin-Nian~Wang,~Phys.~Rev.~Lett~{\bf 96},~172302~(2006).
\bibitem{prc73} A.~Majumder and Xin-Nian-Wang,~Phys.~Rev.~C~{\bf 73},~051901~(2006).
\bibitem{npa767} I.~M.~Dremin,~Nucl.~Phys.~A~{\bf 767},~233~(2006).


\bibitem{prd39} M.~C.Chu and T.~Matsui, Phys.~Rev.~D~{\bf 39},~1892 (1989).

\bibitem{prd68} P.~Romatschke and M.~Strickland Phys. Rev. D {\bf 68}, 036004 (2003).
\bibitem{prl100} M.~Martiez and M.~Strickland, Phys.~Rev.~Lett. {\bf 100},~102301 (2008).
\bibitem{prc78} L.~Bhattacharya and P.~Roy, Phys.~Rev C {\bf 78}, 064904 (2008).
\bibitem{prc79} L.~Bhattacharya and P.~Roy, Phys.~Rev C {\bf 79}, 054910 (2009).
\bibitem{prc81} L.~Bhattacharya and P.~Roy, Phys.~Rev C {\bf 81}, 054904 (2010).
\bibitem{prc83} P.~Roy and A.~K.~Dutt-Mazumder, Phys.~Rev. C {\bf 83}, 044904 (2011).
\bibitem{prc84} M.~Mandal,~L.~Bhattacharya and P.~Roy, Phys.~Rev.~C {\bf 84}, 044910 (2011).

\bibitem{prd70} P.~Romatschke and M.~Strickland Phys. Rev. D {\bf 70}, 116006 (2004).

\bibitem{prd62} St.~Mrowczynski and M.~H.~Thoma, Phys.~Rev. D {\bf 62},~036011 (2000).

\bibitem{ichimaru} S.~Ichimaru, Basic Principles of Plasma Physics (W.~A.~Benjamin, New York, 1973).



\bibitem{ppp} N.~A.~Krall and A.~W.~Trivilpiece, Principles of Plasma Physics (New York: McGraw-Hill). 
\bibitem{Landau} L.~D.~Landau, E.~M.~Lifshitz, Fluid Mechanics, 2nd edition.
\bibitem{ppcf} W.~J.~Miloch, Plasma Phys. Control. Fusion {\bf 52}, (2010) 124004.

\bibitem{prd73} B. Schenke, M. Strickland, C. Greiner, and M. H. Thoma,
Phys. Rev. {\bf D 73}, 125004 (2006).


\bibitem{MM1} M. Mandal and P. Roy, in preparation.  

\end{thebibliography}
\end{document}